# Metered reagent injection into microfluidic continuous flow sampling for conductimetric ocean dissolved inorganic carbon sensing.

**Mark Tweedie[1], Antonin Macquart[2], Joao Almeida[3], Brian Ward[3], Paul Maguire[1].**

[1] NIBEC, Ulster University, Belfast, Northern Ireland.
[2] ams AG, Zurich, Switzerland.
[3] AirSea Laboratory, Ryan Institute & School of Physics, National University of Ireland, Galway

Email: m.tweedie@ulster.ac.uk

**Abstract**

Continuous and autonomous measurement of total dissolved inorganic carbon ($TCO_2$) in the oceans is critical for modelling important climate change factors such as ocean uptake of atmospheric $CO_2$ and ocean acidification. Miniaturised chemical analysis systems are therefore required which are small enough for integration into the existing Argo ocean float network for long-term unattended depth profiling of dissolved $CO_2$ with the accuracy of laboratory bench analysers. A microfluidic conductivity-based approach offers the potential for such miniaturisation. Reagent payload for >3 yr operation is a critical parameter. The precise injection of acid into sample, liberating $CO_2$ from seawater, is addressed here. Laser etched microfluidic snake channel restrictors and asymmetric Y meters were fabricated to adjust the metering ratio between seawater and acid simulants. Laser etching conditions were varied to create a range of channel dimensions down to ~75 µm. Channel flow versus pressure measurements were used to determine hydrodynamic resistances which were compared with finite element simulations using a range of cross-section profiles and areas. Microfluidic metering circuits were constructed from variable resistance snake channels and dimensionally symmetric or asymmetric Y-junctions. Sample to acid volume ratios (meter ratio) up to 100:1 have been achieved with 300 µm wide snake channel for lengths > 1m. At the highest pattern resolution, this would require a footprint of > 600 $mm^2$ (6 $x10^{-4}$ $m^2$). Circuits based solely on asymmetric Y-junctions gave meter ratios up to 16:1 with a footprint cost of < 40 $mm^2$ and precision values of ~0.2%. Further design and fabrication refinements will be required to ensure the structural and dimensional integrity of such small channels in future integration of metering units into full $TCO_2$ analysis microfluidic circuits.

## 1. Introduction

Measurement of total dissolved $CO_2$ ($TCO_2$) content in seawater by miniaturised sensors in deep sea floats is set to become important for long term oceanic monitoring, as the oceans capture the increasing amounts of $CO_2$ released by fossil fuel burning [1]. Such measurements will feed into climate change modelling, to improve the accuracy of future global temperature rise predictions as monitored by the UN [2], and sea level rise prediction [3]. New microfluidic lab on chip devices for $TCO_2$ measurement must be developed for long-term autonomous deployment in a deep ocean environment. This poses severe challenges with regard to design, fabrication and testing. Oceanic studies of $TCO_2$ content in seawater have traditionally used sample collection by research vessels and laboratory analysis. This is slow, expensive and has provided only sparse coverage, whereas continuous remote monitoring across the world's oceans and to depths of up to 3km are required. Several methods have been demonstrated for field analysis of $TCO_2$, such as conductivity change [4 - 6], spectrophotometry [7 - 10], IR absorption analysis [11 - 14], gas chromatography [15], and membrane inlet mass spectroscopy [16 - 18]. Partial pressure of surface $CO_2$, $pCO_2$, has been measured from buoys, floats, and ships [19 - 21]. Microfluidics based conductivity and temperature sensors have been trialled in the ocean [25], with other devices also investigated for oceanic nitrate/nitrite and phosphate/phosphorus detection [28 - 31], ammonium ion measurement [32], manganese sensing [33] ocean acidification [34], biochemical and microbiology applications in similar extreme environments [26, 27].

The Argo network of ~5000 untethered autonomous ocean floats offers the possibility of continuous low cost $TCO_2$ measurements to the required spatial and depth resolution and therefore inspires the development of a suitable $TCO_2$ microfluidic platform which could meet the stringent constraints on size and power, as well as reliability in the harsh ocean environment. Argo floats are currently deployed to measure pressure, temperature and salinity using real time (instantaneous) sampling and measurement as they ascend to the surface from their target depth of 2 – 3 km [39]. The floats then transmit the data to satellite, descends to a drift depth of 1500 m for 9 days, before the next descent to target depth. The 10-day cycle repeats continuously, with each untethered float having a projected lifetime of 3 - 5 years. $TCO_2$ measurements using laboratory prototypes or in short term ocean deployment have achieved the $\leq 0.2\%$ precision thought to be necessary for ocean characterisation [5, 9, 10, 37]. However microfluidic implementations which meet volume, power, cost and reliability constraints struggle to reach the required accuracy. $TCO_2$ determination by conductivity measurement was originally proposed by Hall and Aller [4] and remains the approach with the greatest potential for low-cost and low-power miniaturisation compared to optical or mass spectrometry. However, with conductivity measurement, noise rejection is a much more demanding challenge.

Total dissolved $CO_2$ ($TCO_2$) in seawater exists mainly in the form of $CO_3^{2-}$ (carbonate) and $HCO_3^-$ (bicarbonate) ions, with small amounts of carbonic acid, and dissolved free $CO_2$ molecules. Acidification of seawater to reduce the pH (<4) converts the $TCO_2$ forms to aqueous $CO_2$ which can be separated from seawater by diffusing across a membrane into a receiving solution of NaOH. The $CO_2$ then reacts with NaOH to form $CO_3^{2-}$ (pH $\geq$ 12) via the dominant reaction $CO_2 + 2OH^- \rightarrow CO_3^{2-} + H_2O$, Fig. 1 [4]. Under appropriate conditions, the replacement of $OH^-$ with the lower mobility $CO_3^{2-}$ ions provides a linear relationship between conductivity and $TCO_2$ concentration. A microfluidic $TCO_2$ analysis system would therefore consist of several integrated functional units to achieve (i) acid injection and mixing in seawater for pH reduction, (ii) membrane liquid – liquid exchange of aqueous $CO_2$ and (iii) conductivity measurement in a high concentration alkaline environment. The time taken for $CO_2$ membrane exchange via diffusion prevents instantaneous measurement as the float ascends or descends and hence the only option is to collect seawater samples in microfluidic storage cells during ascent/descent for subsequent analysis while the float is parked at 1500 m for 9 days. Multi-sample membrane-based analysis poses new challenges for microfluidics fabrication including a requirement for high integrity multi-channel and multi-level structures with robust, long-term and chemically-resilient bonding. In recent work we have demonstrated suitable PMMA bonding processes for long term multi-layer and multi-channel operation [42], PDMS $CO_2$ membrane sealing within a PMMA manifold [43] and microfluidic $CO_2$ separation and conductivity measurement [44]. The requirement for sample storage plus the reagent (acid, NaOH) payload for measurement and flushing represents the primary contribution to system volume and maximum limits below ~1L will be necessary. With P profiles (up to 150) per float life and N samples per profile (up to 100), reagent usage must ultimately be reduced to the microlitre per sample scale.

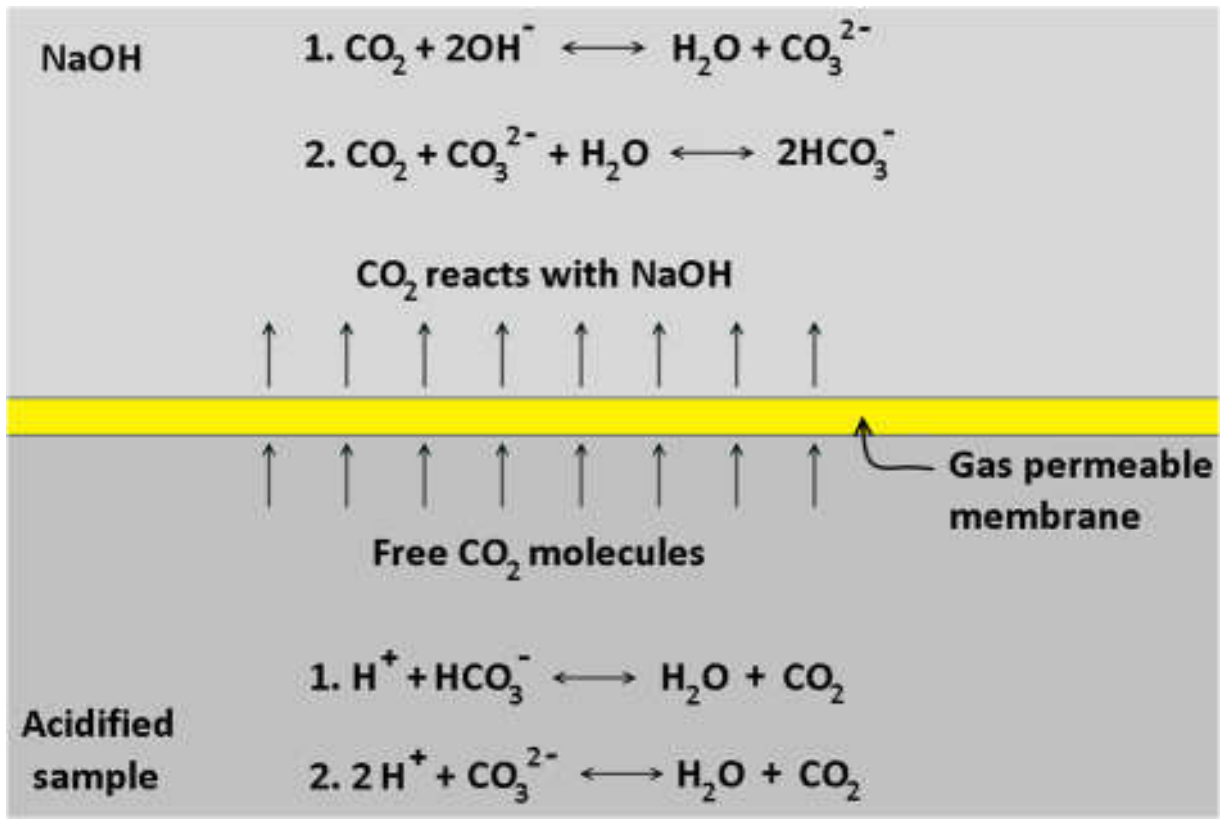

Fig. 1. Gas diffusion from acidified sample into NaOH, for $TCO_2$ measurement.

This work investigates one such unit, the injection of precise acid quantities into seawater. Reduction in seawater pH from its normal level (~8) to $\leq 4$ is possible by adding a minimum quantity of high strength acid. To avoid the necessity of a separate acid pump which incurs a significant power and volume cost, simultaneous acid injection during the seawater sampling stage offers a possible solution. The controlled additional of acid is determined by the metering ratio (MR) of the microfluidic circuit where MR is the sample to acid volume ratio ($V_S/V_A$). The total acid payload volume is therefore $PxN(V_S/MR)$ and maximising MR without incurring sample volume or other fabrication penalties is the objective. For a monoprotic acid, A, the maximum MR at which all $TCO_2$ is converted to $CO_2$ is $[A]/([TCO_2]+[CO_3^-])$, Fig. 1, which is ~300[A] for an upper $[TCO_2]$ limit in seawater of < 3 mM and a carbonate fraction of ~10% at pH8. The addition of acid dilutes the sample and hence affects measurement precision, given by $P = (1 + MR)^{-1}$. To achieve a desired precision, P, the conditions to be met are

$$MR \geq \frac{1-P}{P}, \; [A] \geq \frac{[TCO_2]+[CO_3^-]}{P}$$

A precision of 0.1% would require an acid concentration $[A] \geq 3$M and MR ~1000 which is not realistic. The alternative therefore is to include the dilution factor within the device calibration. In this case the precision is now dependent on MR fluctuations during operation due to, for example, temperature and pump variability. The maximum allowed fluctuation in MR from the original calibrated value is shown in Fig. 2, for a specified precision of 0.1%. Since the resistance of acid and seawater channels depends on dynamic fluid viscosity, the MR temperature dependence needs to be considered over the likely ocean temperature variation from surface to 2 km depth. This variation is typically 10 °C but can reach up to 20 °C. The change in MR with temperature is calculated for two rectangular cross-section channels, Fig. 3, from the standard analytical approximation, (1).

$$R_H = \frac{12\mu L}{wh^3\left(1-\frac{h}{w}\left(\frac{192}{\pi^5}\sum_{n=1}^{\infty}\frac{1}{(2n-1)^5}\tanh\left(\frac{(2n-1)\pi w}{2h}\right)\right)\right)} \quad (1)$$

where $R_H$ is the hydrodynamic resistance of a rectangular channel of length, L, width, w, and height, h, carrying a fluid of dynamic viscosity, μ. The nominal MR, given by $R_{H(acid)}/R_{H(SW)}$, is dependent on the ratio of acid (HCl) to seawater dynamic viscosity which reduces with depth, as the sea temperature falls, Fig. 3. For an expected temperature at depth of 5 °C and a maximum surface temperature of 25 °C, assuming 20% HCl concentration and an ocean salinity of 0.035 kg $kg^{-1}$, the change in MR is ~8%. To limit the effect of such a MR variation on $TCO_2$ measurement precision implies a minimum MR value of ~75, Fig. 2. With lower HCl concentration, the temperature dependence is attenuated. While it would be possible to factor the temperature dependence into the device calibration, this would require simultaneous temperature measurement at each depth sampling point.

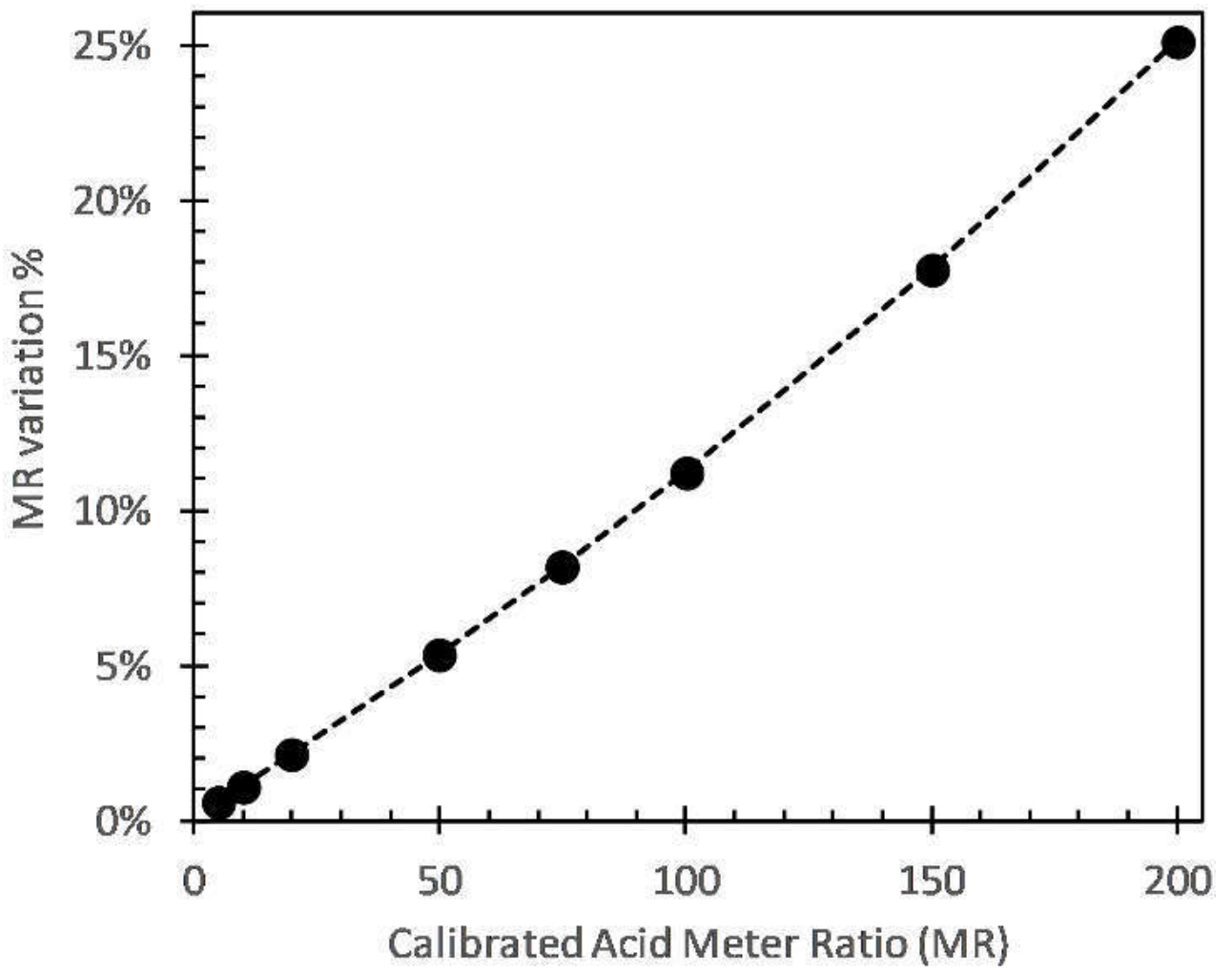

Fig. 2. Effect of fluctuation in meter ratio (MR). The maximum allowed variation versus the calibrated MR to maintain precision ≤ 0.1%.

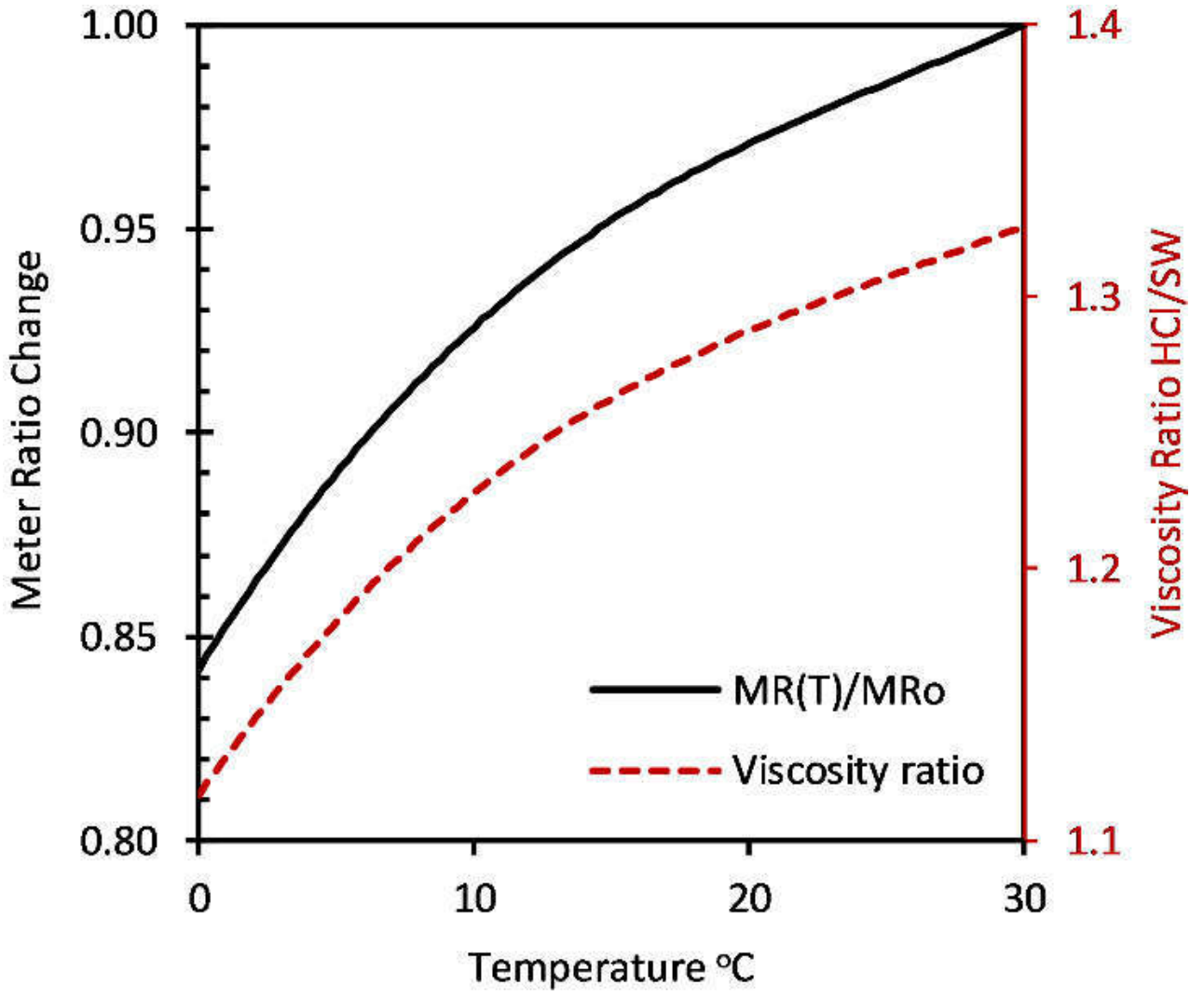

Fig. 3. Calculated change in meter ratio (MR) and viscosity ratio of HCl (20%) and seawater, with salinity of 0.035 kg $kg^{-1}$. The calculations were based on equation 1, with a range of nominal MR values from 15 to 120. This represents the range of experimental MR values achievable using rectangular cross-section channels fabricated by precision milling.

In this paper we report the fabrication and characterisation of microfluidic structures suitable for single pump use in a continuous flow operation where metering is based on the channel resistance difference between two arms of a microfluidic junction. We implement various MR values via optimisation of channel cross-section area (CSA) or length. The latter involves uniform CSA which facilitates ease of fabrication at the expense of volume while the former involves differential CSA and presents a challenge for fabrication due factors such as channel height loss and deformation which are inevitable outcomes of thermoplastic bonding. Microfluidic performance has a strong dependence on the minimum achievable channel height and width values and their associated tolerances. While we have obtained sub-100 μm using precision milling, the need for long tightly spaced channels and variable heights and widths within a given pattern have proven to be problematic with this method. Instead we use laser etching to define channels, due to the pattern flexibility and higher resolution. However, laser etching, especially at the finest resolution, results in a loss of the traditional rectangular cross-section and hence the resultant channel resistance is no longer predictable. Along with pressure – flow and meter ratio measurements, we also report a number of finite element simulations with variable cross-section profiles used to estimate channel resistance.

## 2. Experimental

Microfluidic channels were engraved in cast PMMA (10 mm thickness) using a $CO_2$ laser, (Universal Laser Systems VLS2.30, 25W source power, wavelength 10.6 μm), from AutoCAD patterns. A standard 1.5" focal length lens was used for rastered snake channels, with lengths of 330, 663, 994 and 1320 mm, a typical example being shown in Fig. 4. The snake channels consisted of 30 mm straight sections with semi-circle turns at each end. All snake channels were rastered with settings of 100% power, 25% speed (~ 0.3 m/s), and AutoCAD drawing linewidth of 0.09 mm. A HPDFO (High Power Density Focussing Optics) lens was used for improved optical definition for fabrication of Y meters, which required finer linewidths for narrow restrictions in a section of the acid input line. Fluidic connections were then attached via milled and tapped ¼-28 ports with 1mm through holes. PMMA bonding was achieved via $CHCl_3$ solvent vapour treatment and subsequent thermal treatment (50°C, ≥ 20 h), to drive off residual chloroform. Asymmetric Y meters were fabricated with raster mode channels of nominal channel width 0.5 mm, except for a thin 10 mm long vector mode restrictive channel between the acid line input and the Y-junction. Laser vector mode operation gives a smoother line with straight edges in the Y direction. Four Y meters with different restrictive channel cross-sections were fabricated, Fig. 5, with the Y junction axis at 45° or 225° rotation.

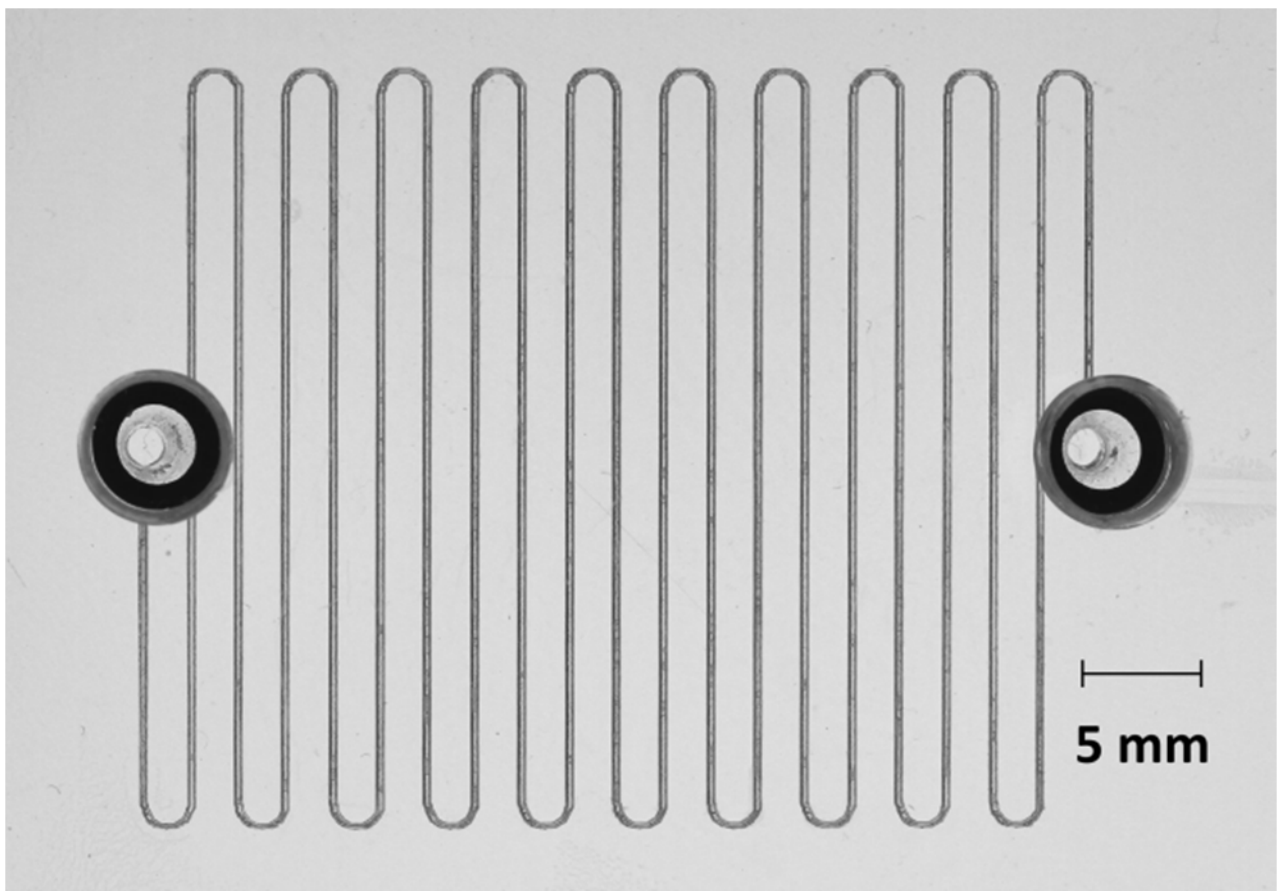

Fig. 4. Snake channel, 663 mm length, with ¼-28 tapped ports.

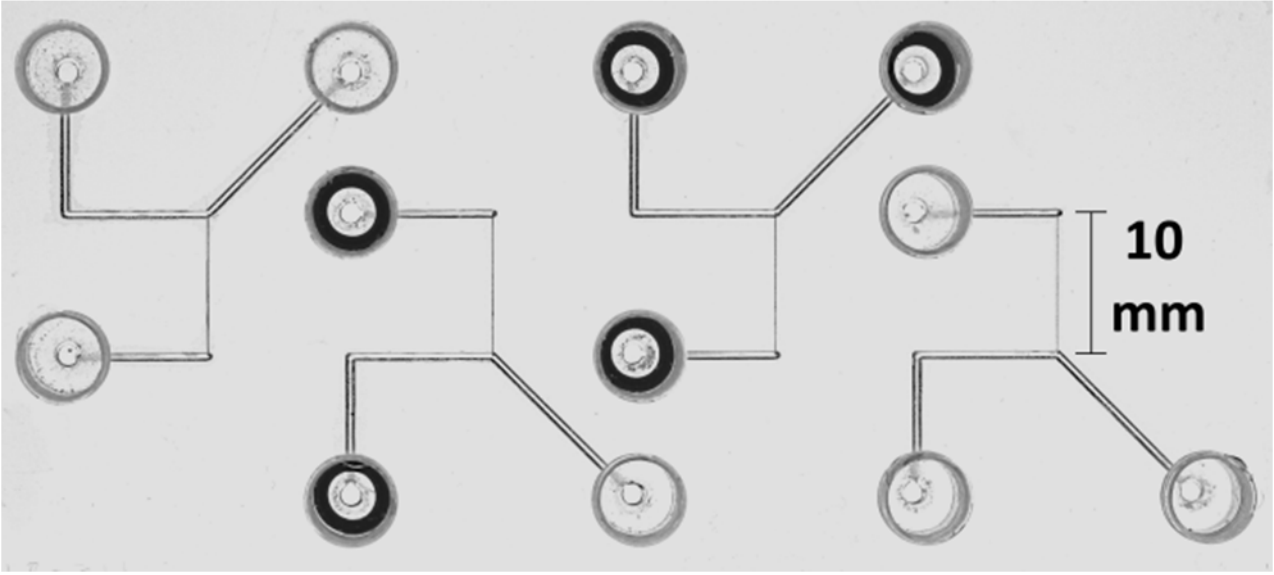

Fig. 5. Asymmetric Y meters, with ¼-28 tapped ports. Each Y meter has a different resistance in one input, via use of vector writing for narrower channels.

Representative optical cross-sections of raster and vector-written channels in PMMA are shown in Fig. 6 and Fig. 7, respectively. The raster channels have a flatter base, because they are formed from at least 2 scan lines, while the vector channels are formed from one scan line, and so tend to V-shapes. The rastered example is shown without a bonded lid, while the vectored channels are shown after bonding to enclose them. Maximum channel width and height dimensions of the enclosed vector channels were obtained from image analysis (ImageJ), for a constant laser speed of ~ 0.3 $ms^{-1}$, Fig. 8. The cross-sectional area was determined using the ImageJ tracing function and area versus laser power is observed to be linear. Area values obtained using a trapezoid or triangular profile showed average errors of 2% (±13%) and 16% (±13%) respectively. The rastered snake channel cross-sectional area was 7.35 $x10^{-8}$ $m^2$, which is > 5x the largest vector area, and measured volumes ranged from ~ 18.5 μL to 73.5 μL.

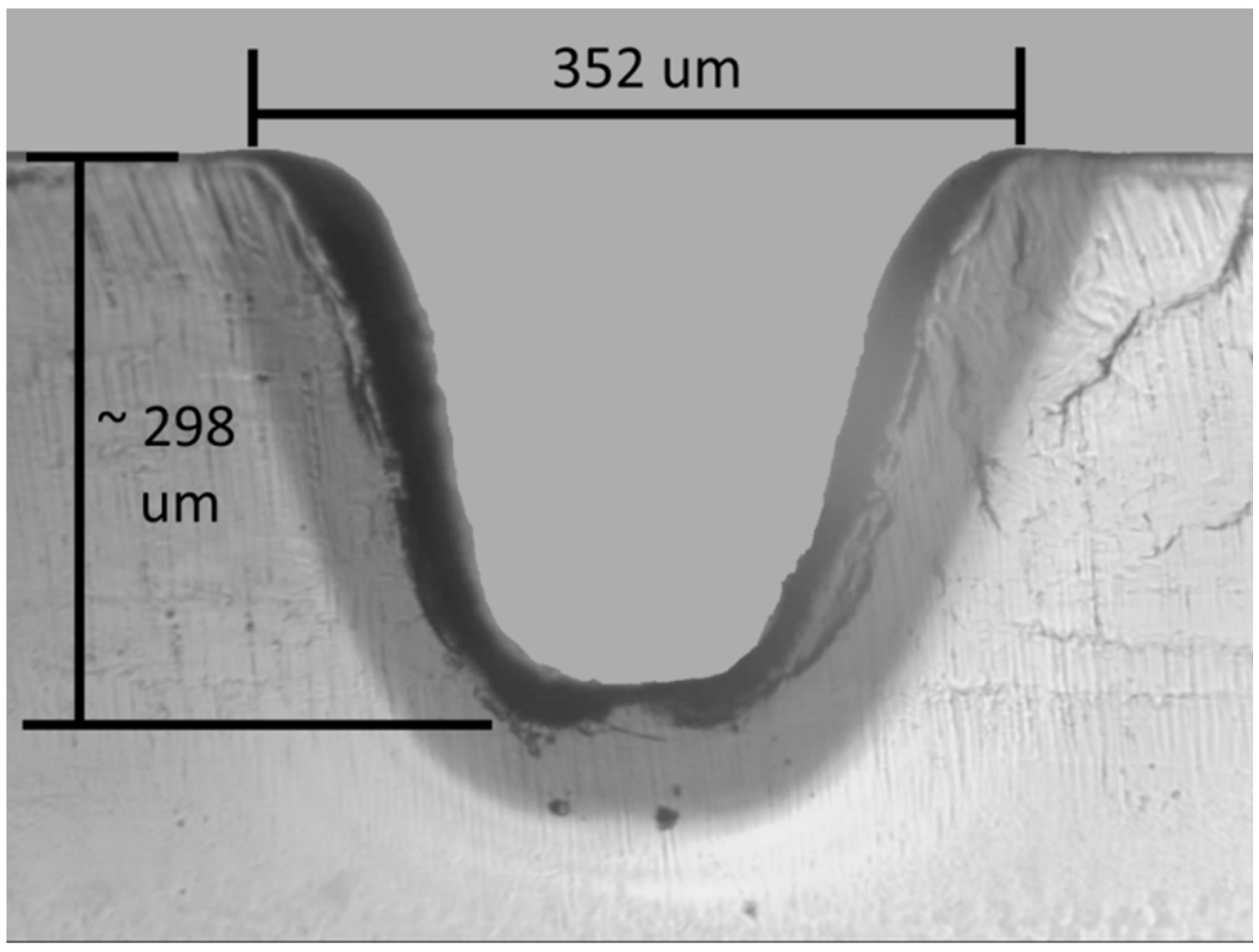

Fig. 6. Rastered channel optical cross section, fabricated via a 1.5" lens.

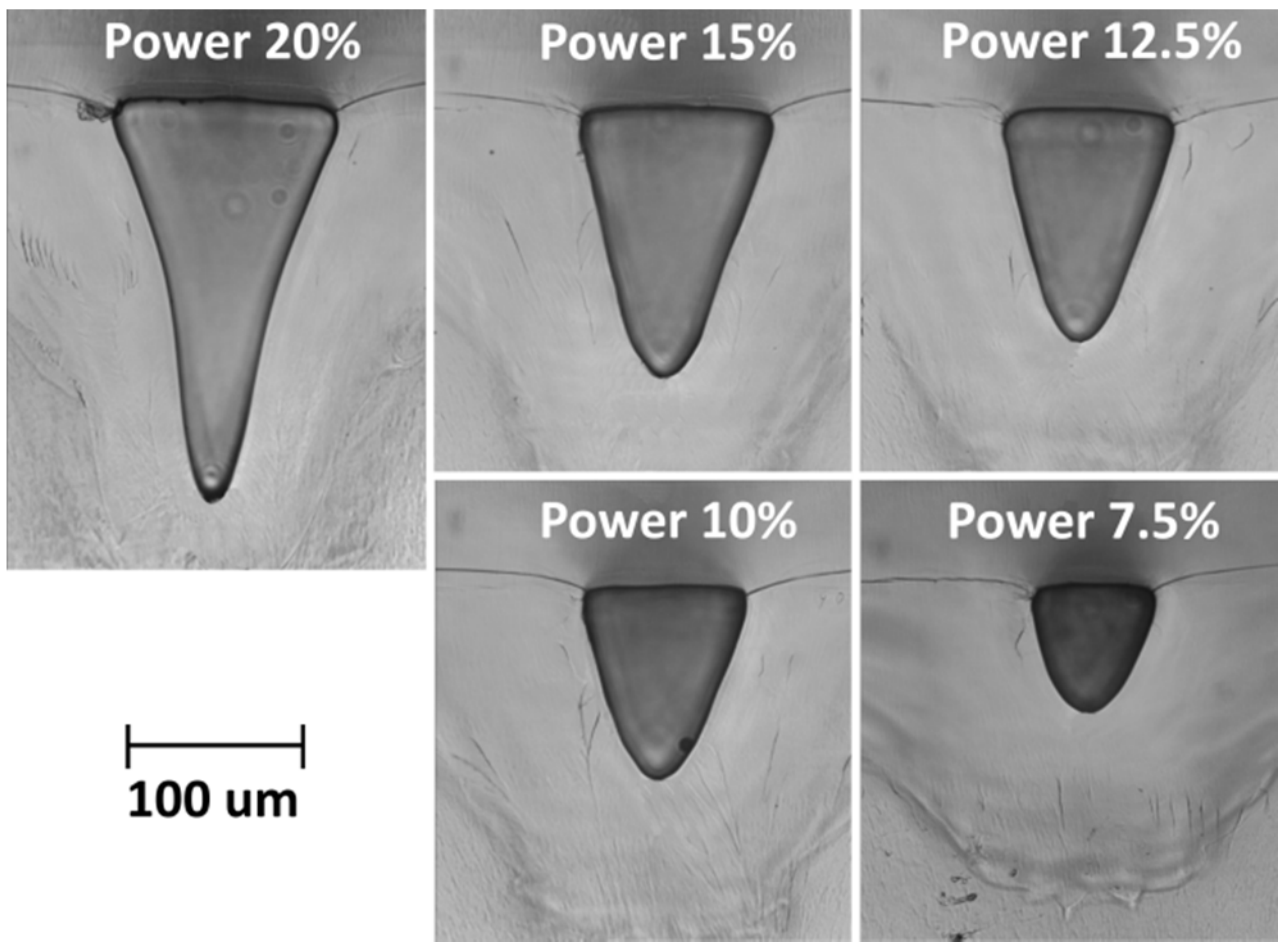

Fig. 7. Vector written channel optical cross section (using HPDFO lens), post bonding, as a function of laser power (%). The maximum power is 25 W.

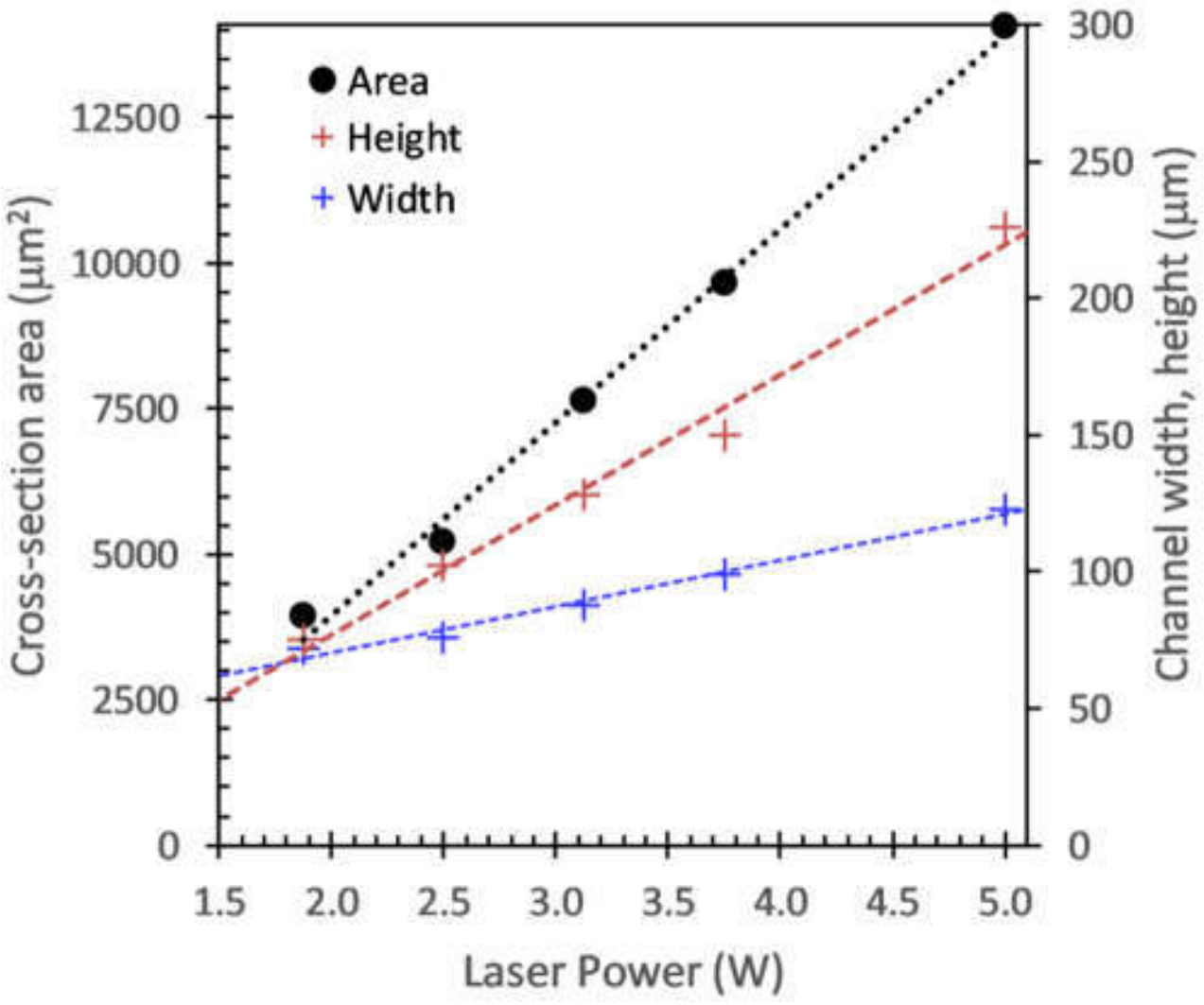

Fig. 8. Dimensions of laser etched channels using vector mode with a HPDFO lens versus laser power. Maximum channel width (W) and height (H) dimensions were obtained from microscopy and cross-sectional area obtained from autotracing.

Metering ratios were determined using KCl standard solutions (Hanna Instruments) with high (1350 – 1550 μS/cm) and low (84 – 95 μS/cm) conductivity and DI water (1 – 2 μS/cm). Conductivity values were confirmed using a temperature-compensated Metrohm 712 probe meter. An Elveflow AF1-P1600 pressure generator was used to drive fluid through microfluidic devices. Two experimental schematics are shown in Figs. 9 and 10, for snake channels and Y meters, respectively. Elveflow flow meters with maximum rates of 1mL $min^{-1}$ and 5 mL $min^{-1}$ under ESI software control were used to measure flow rates for known positive pressures and the hydrodynamic resistance, $R_h$, obtained from the inverse slope of flow rate versus pressure. For snake cells, high conductivity (~ 1500 μS/cm) and DI water solutions were simultaneously pumped using cells of different lengths positioned in each arm of the fluidic set up, Fig. 9. The outputs were combined at an equal T-piece, and the resultant conductivity measured in a 10 mL cell containing a temperature compensated 712 Metrohm conductometer probe.

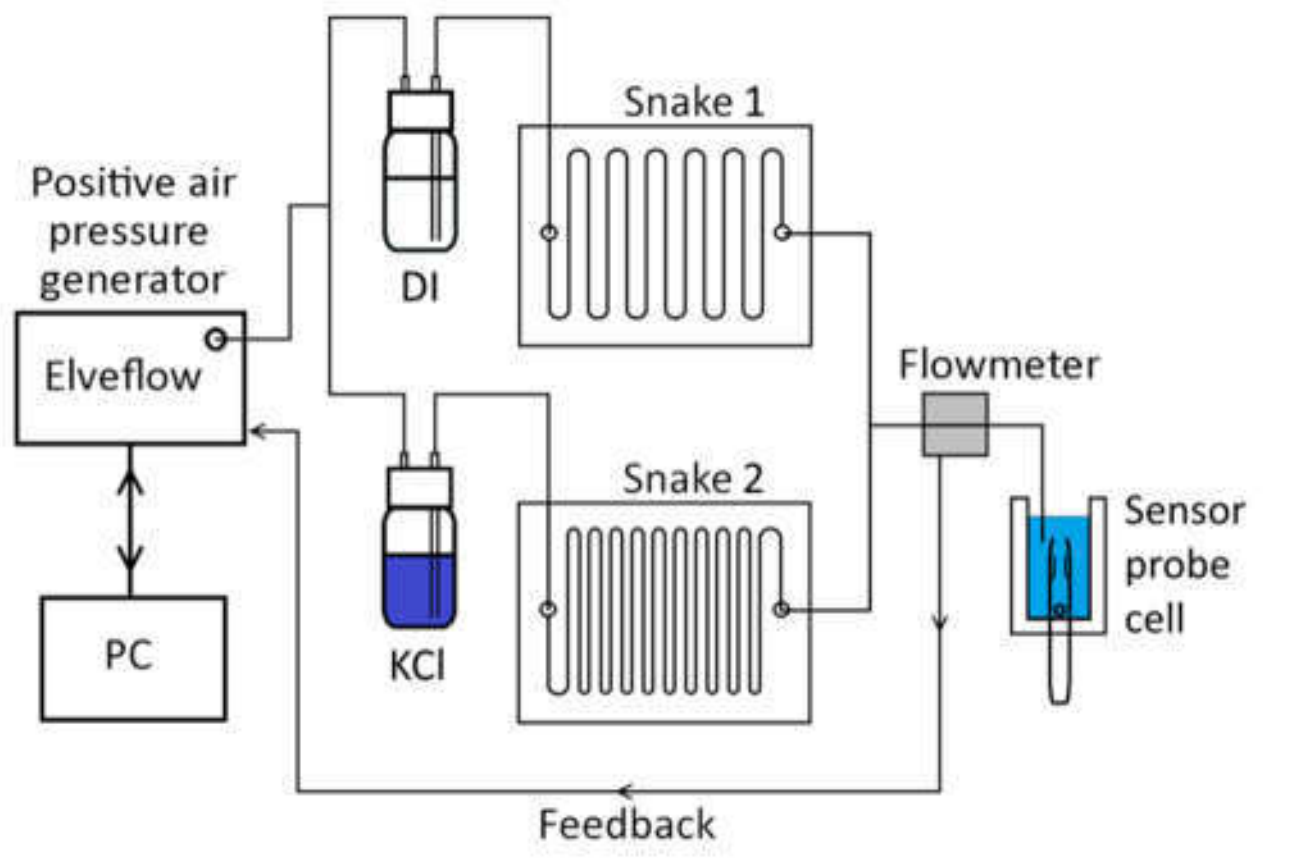

Fig. 9. Elveflow positive pressure schematic for snake cell meter ratio testing, with flow sensor feedback (PC not shown).

The metering ratio is defined as the volume ratio of the two solutions (MR = $V_1 / V_2$), where $V_1$ represents the seawater equivalent solution and $V_2$ represents the high conductivity acid solution. The final total volume is $V_T = V_1 + V_2$. By closing the $V_2$ channel, the value of $V_1$ can be determined. Alternatively, by assuming the same relationship between ion concentration and conductivity for all solutions, then $\sigma_T V_T = \sigma_1 V_1 + \sigma_2 V_2$ which can be rearranged to give

$$MR = \frac{V_1}{V_2} = \frac{\sigma_2 - \sigma_T}{\sigma_T - \sigma_1} \qquad (2)$$

A similar set-up was used to measure the metering performance of 3 different Y meters, Fig. 10.

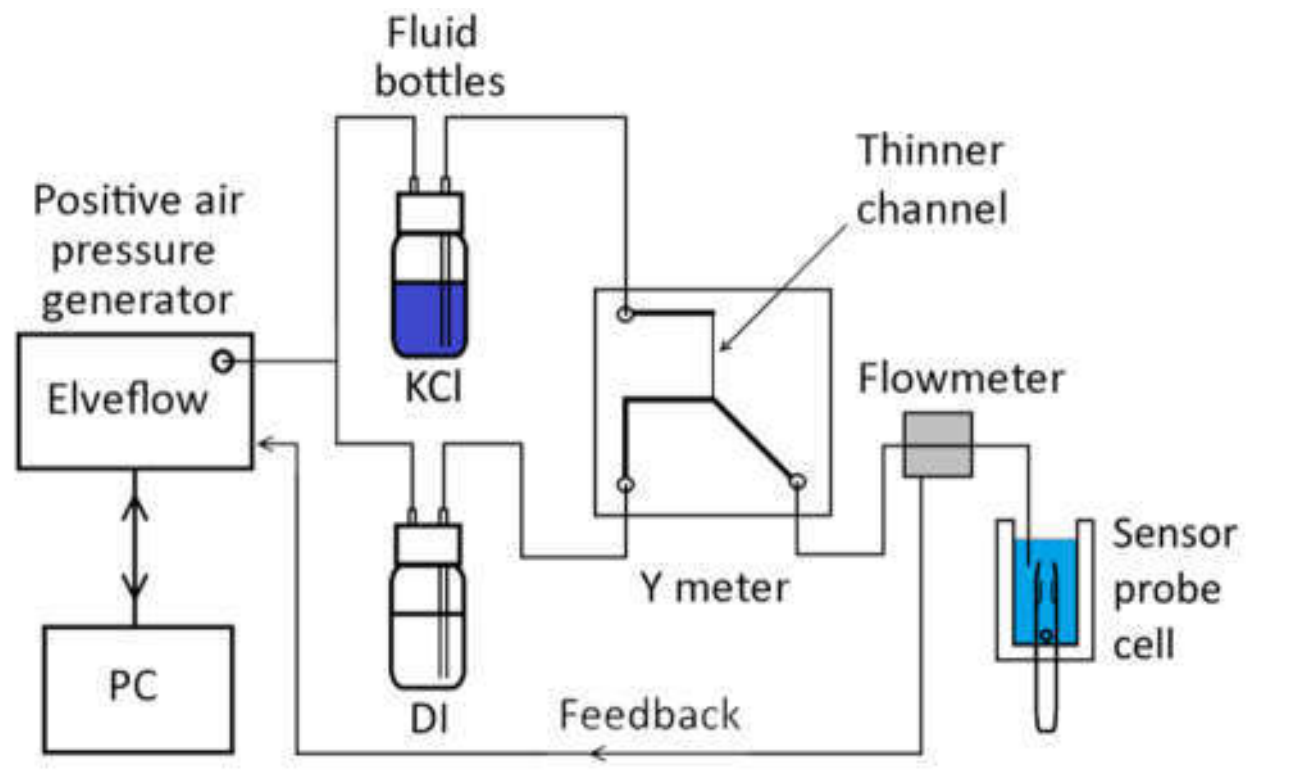

Fig. 10. Elveflow positive pressure schematic with flow sensor feedback for Y meter measurements.

Tests were also performed under negative pressure, where the two fluids are simultaneously drawn through the devices, so representing the preferred configuration for an ocean deployed device for acid/seawater mixing. Cetoni NeMySys syringe pumps were used with precision glass-bodied Setonic syringes, as in Fig. 11. Here syringe 1 pulls fluid from both high and low conductivity solutions through the Y meter test device, and then dispenses 8 mL of the mixed solution into the conductometer probe cell. Aluminised DaklaPack 1L spoutbags, specified as suitable for gasless liquids, were used to hold the fluids. The caps were fitted with Diba PTFE 2-way valves, and silicone applied to prevent fluid leakage. Syringe 2 in Fig. 11 takes fluid from the low refill bag and replenishes the low solution bag to its initial level, immediately after each mix sample is taken. Refill of the high solution bag is not necessary since the low:high mix ratios is >10:1.

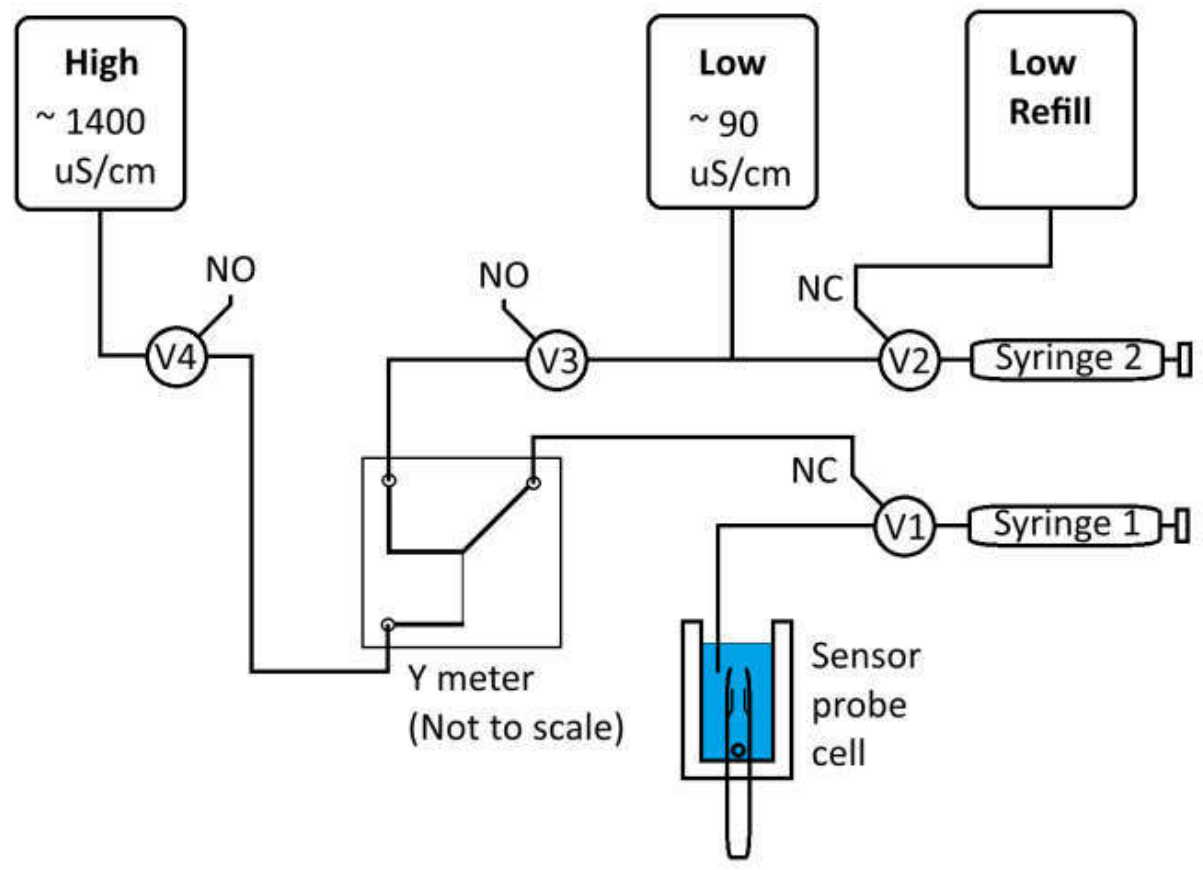

Fig. 11. Cetoni NeMeSys syringe negative pressure schematic (PC not shown), for Y meter testing, with fluid in bags.

## 3. Simulation

The theoretical hydrodynamic resistance, $R_H$, of a rectangular cross-section channel depends on width, w, and height, h, as in equation (1). For circular cross-section of radius R, it depends on $R^{-4}$. The high sensitivity of $R_H$ to dimension and shape means that routine methods of estimating $R_H$ for laser etched cross-sections are not available. Finite element modelling of fluid flow in channels of various cross-sectional shapes and dimensions was carried out using Comsol Multiphysics software (v5.3). Snake channel hydrodynamic resistance was determined by modelling one complete straight portion of 30 mm length, and 1 turn of diameter 2 mm, then multiplying the result by the number of turns. Channel cross-sections were modelled using a trapezoidal approximation with varying base widths, $W_B$, or using a curved profile obtained from a fit to the imaged cross sections. The rastered channel cross-section is shown in Fig. 12, with a base width of ~ 140 μm estimated from the flat region, and a comparison between trapezoid and smoothed fitted curve profiles is shown in Fig. 13. The maximum channel height and width are 352 μm and 298 μm respectively. From simulation, $R_h$ increases as the base width of the trapezoid or fitted profile is reduced towards a triangular cross-section, with maximum of change of > 200%, Fig. 14. The resistance obtained from the fitted profile was up to 25% lower and less sensitive to base width. Resistances from both profiles are equal when the curved profile base width is ~40 µm less than that of the trapezoid, Fig. 15.

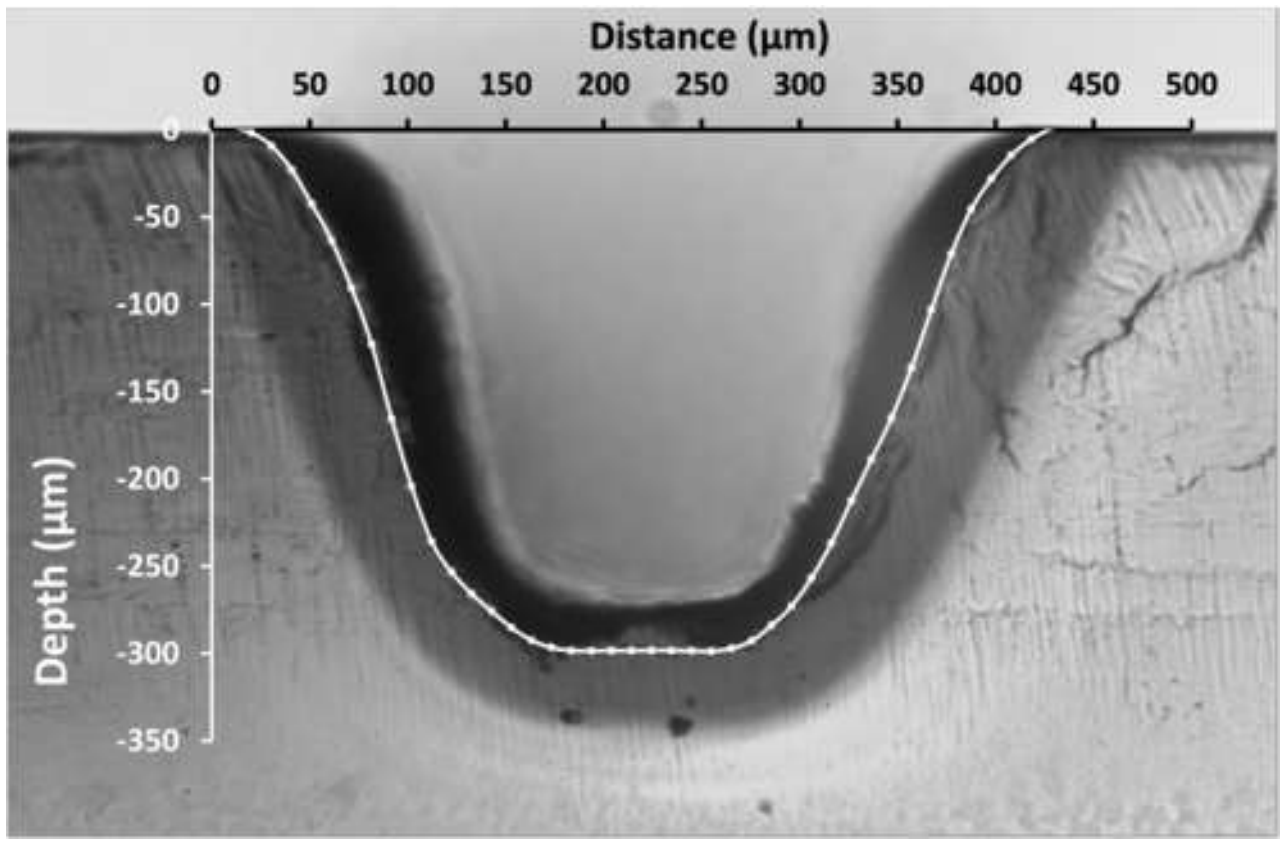

Fig. 12. Outline of rastered channel, superimposed on original channel cross-section.

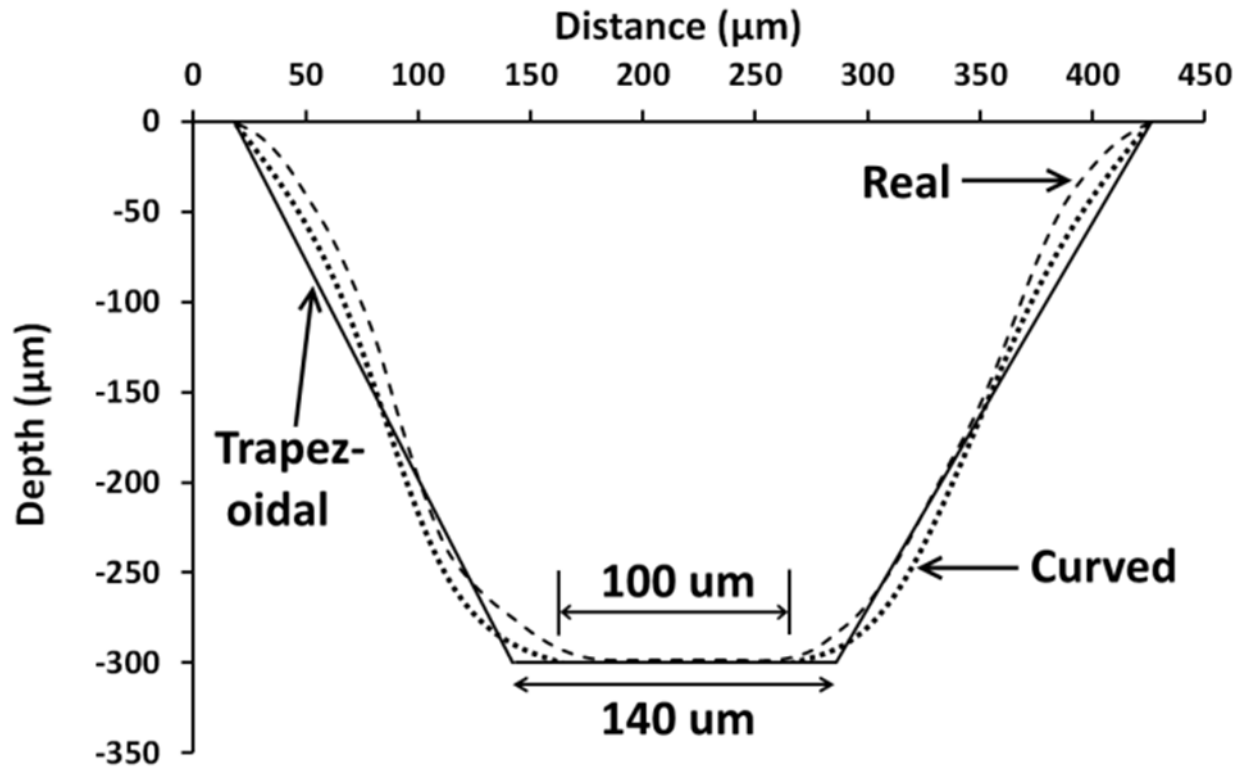

Fig. 13. Approximate trapezoidal, curved and real wall profiles for rastered snake channel.

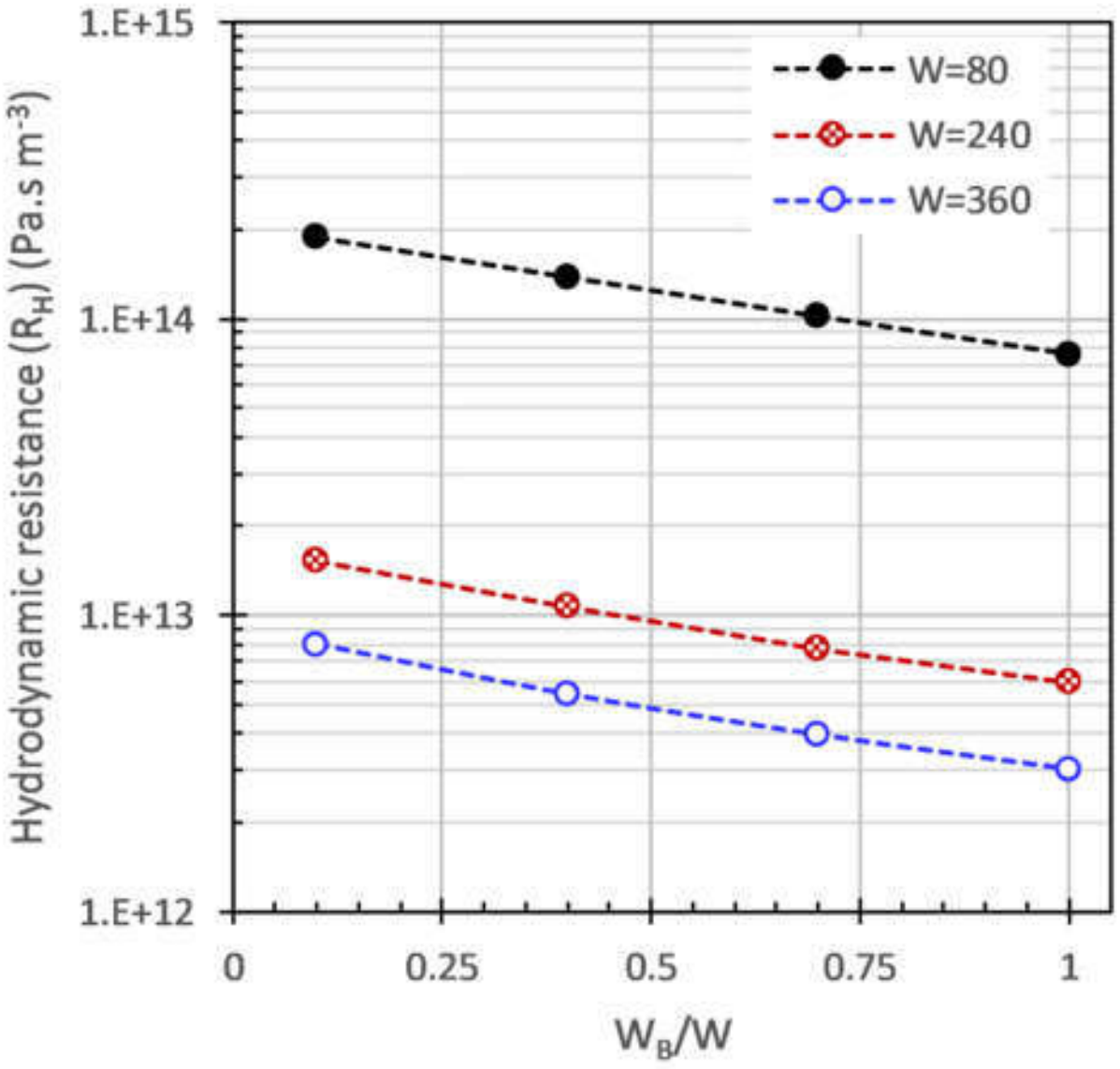

Fig. 14. Impact on $R_H$ of varying base width ratio ($W_B/W$) for top width values, W, from 80 µm to 360 µm. Simulation using fitted profile.

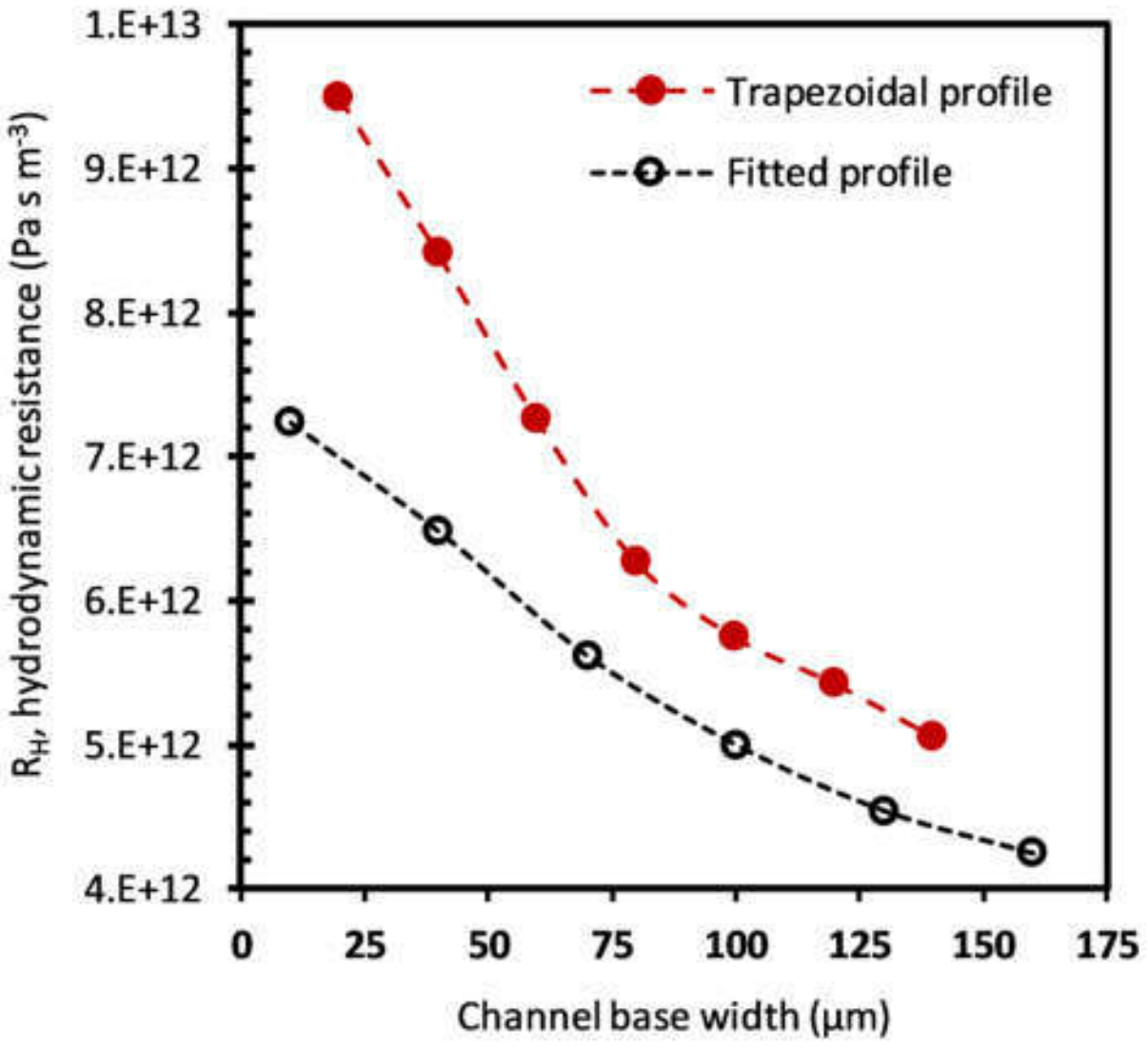

Fig. 15. Simulated $R_H$ versus base width for channels of curved and trapezoidal profiles with 352 µm and 298 µm width and height respectively.

## 4. Results and Discussion

Graphs of flow rate versus differential pressure for each of the 4 snake channels were found to be linear for all devices, an example being shown in Fig. 16. The hydrodynamic resistance, $R_H$, obtained from the inverse gradient is given in Fig. 17 and shows, for constant cross-sectional area, $R_H$ proportional to snake channel length, as expected. Simulated values of $R_H$, using a trapezoidal cross-section with a base width of 140 µm and depth of 298 µm, as obtained from imaging, indicate a 15% reduction compared to the measured values and with a fitted curve cross-section, the simulated value is lower by a further ~10%. One possible reason for this discrepancy, is the reduction of channel height and possible channel deformation observed after PMMA bonding where we have previously observed 15 µm – 25 µm height loss after $CHCl_3$ vapour assisted bonding. [42]. The simulated impact of height loss on $R_H$ is shown in Fig. 18, for the trapezoidal approximation, where the 15% resistance difference is equivalent to 23 µm height reduction. In achieving narrower channels for the high resistance acid line, we observe a change in profile from trapezoidal to almost triangular, with dimensions given in Fig 8.

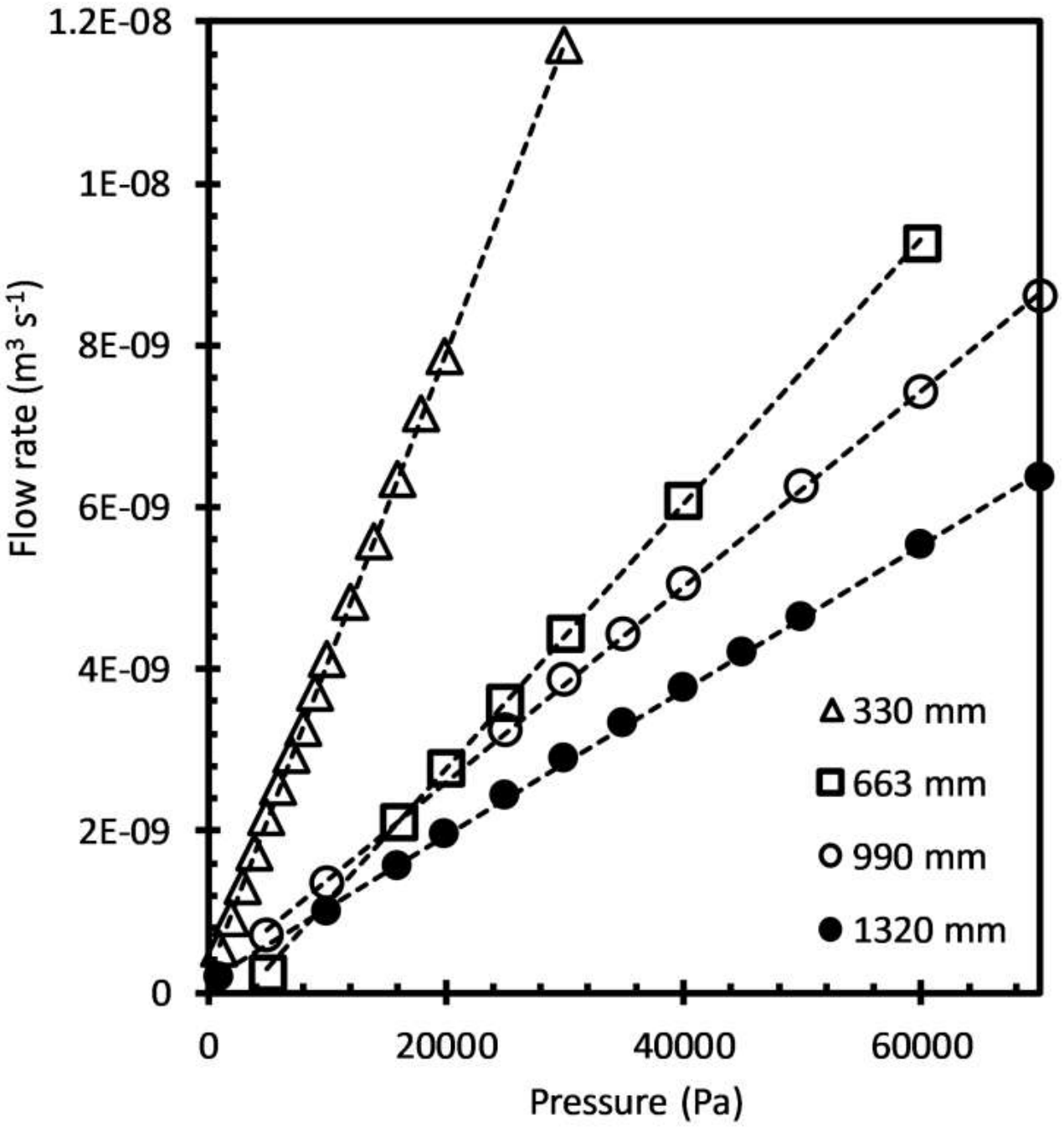

Fig. 16. Flow vs. pressure for snake channels of various lengths with dimension as in Fig. 12. Inverse gradients of fitted lines gives hydrodynamic resistance, $R_H$.

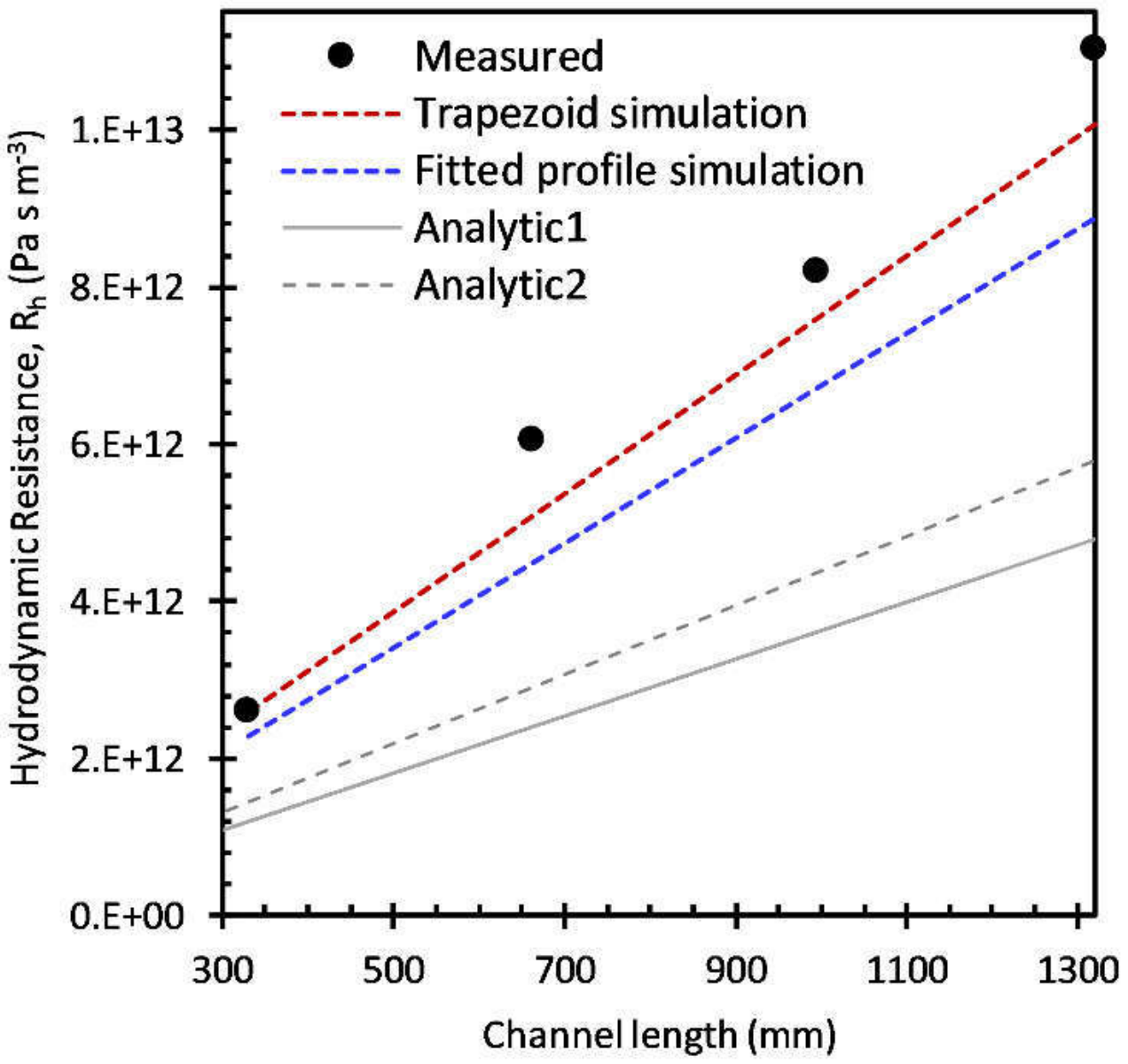

Fig. 17. Measured hydrodynamic resistance, $R_H$, of snake channels versus length compared with simulated values using (i) trapezoidal and (ii) fitted curve cross-sections with a nominal base width of 140 µm and depth of 298 µm. Also shown are analytical calculations (equation 1) for nominal (solid) and reduced heights (dashed).

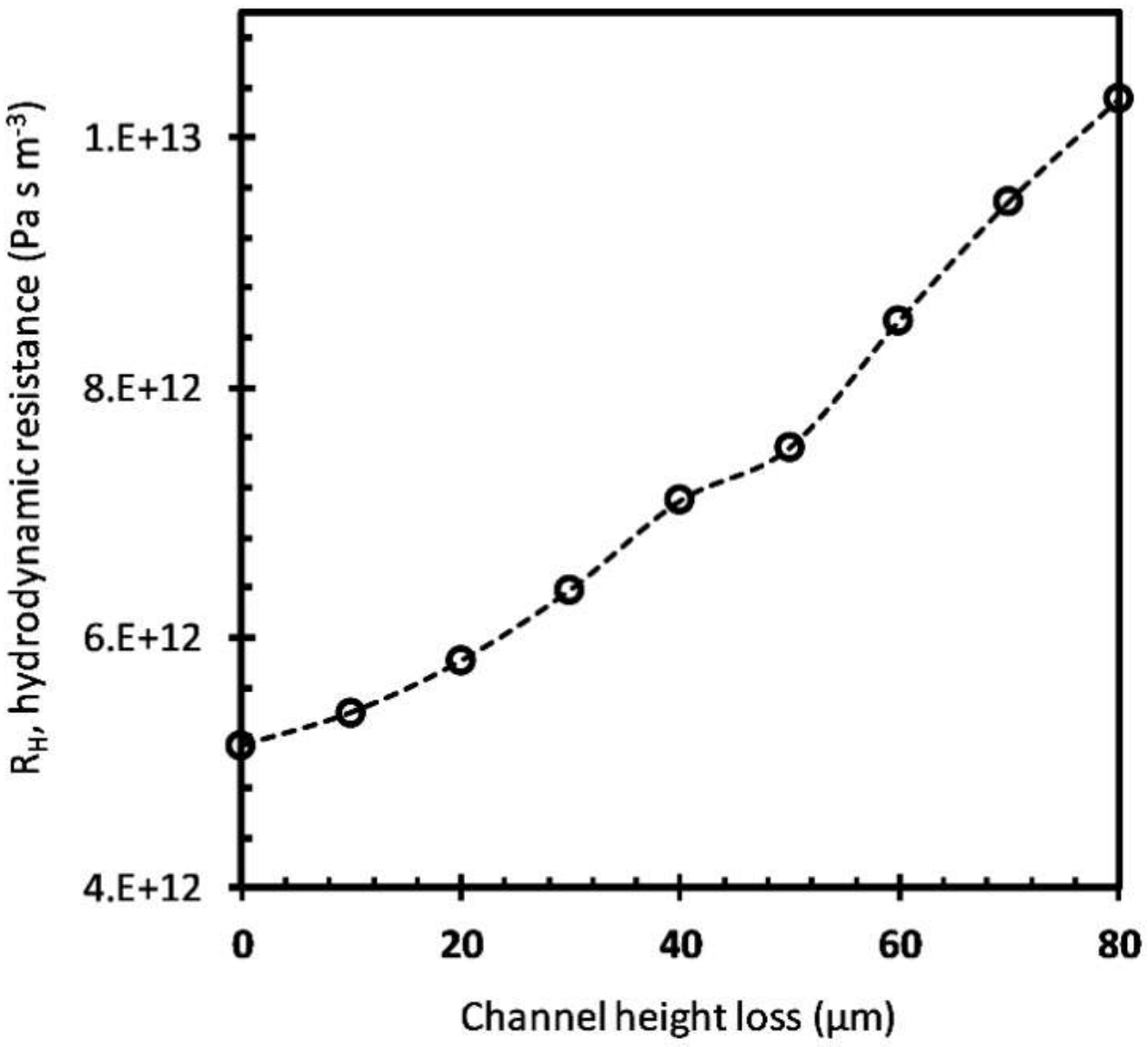

Fig. 18. Simulated $R_h$ versus reduction in channel height for a trapezoidal cross-section of 140 µm and a nominal height of 298 µm.

Metering ratios with snake channels in both high and low conductivity lines were determined from conductivity measurements and compared to those calculated from $R_H$ measurements from the same devices. A 1320 mm snake was inserted in the high conductivity line (acid equivalent), while the low conductivity line was formed from each of the shorter snake channels in turn. The operational positive pressure was 60 mBar. Table 1 lists the measured meter ratios for each of the 3 versions, with a maximum of > 4 for the shortest snake. Measured MR values and those predicted from $R_H$ agree to within ~4%, illustrating that the meter ratios can be predicted with reasonable accuracy from the measured $R_H$ values. These lower MR values may be used where the available valve technology is not specified for long term use with acid concentrations greater than ~ 0.1 M.

| Low conductivity channel length (mm) | 330 | 663 | 994 |
|---|---|---|---|
| Measured mixed solution conductivity, σ (µS/cm) | 285 | 543 | 658 |
| MR (from σ) | 4.29 | 1.78 | 1.29 |
| MR (from $R_h$) | 4.24 | 1.82 | 1.35 |
| MR difference (%) | 1.1% | - 2.3% | - 4.1% |

Table 1. Metering Ratios (MR) for 1320 mm snake in high conductivity line, and shorter channels in low conductivity line, compared to MR's calculated from $R_h$ measurements. MR's (from σ) use a value of 1510 µS/cm for the high conductivity solution.

Meter ratios were also obtained from asymmetric Y-junction devices with one arm containing a short high resistance narrow channel obtained from vector mode laser etching, with dimensions as given in Fig. 8. Measurements were taken using both positive (Elveflow) and negative (Cetoni) pressure driven flow and MR values from 7 to 16 were achieved, in direct relationship to the measured channel resistance, Fig. 19. The small difference between positive and negative pressure measurements in likely due to the differences in external tubing length between both setups. Finally, the incorporation of an asymmetric Y-junction along with a single snake channel, of various lengths, was used to increase the MR above 100, Fig. 20. This illustrates the trade-off between area and MR. The area of the snake, at the highest pattern resolution, would be > 600 $mm^2$ ($6 \times 10^{-4}$ $m^2$) while that of the Y-meter is < 40 $mm^2$. The use of narrow/shallow vector mode channels, which have shown up to x35 increase in resistance per unit length, would reduce the area requirement by up to a factor of 10. However, the integrity of long and shallow channels has yet to be demonstrated within the constraints of a full fabrication process, as illustrated here [42].

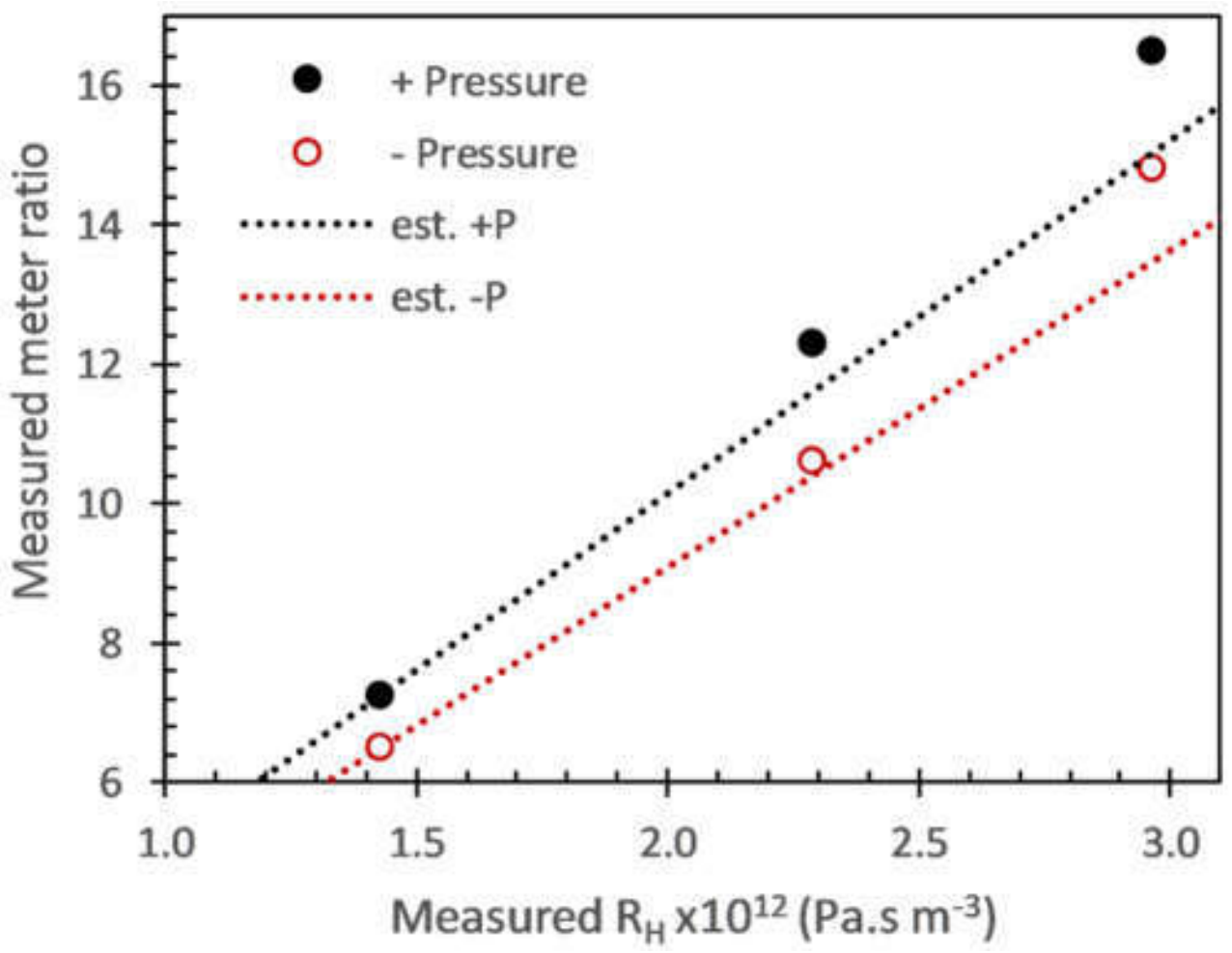

Fig. 19. Single high resistance arm Y-meters: measured Y meter ratios under positive pressure (Elveflow) and negative pressure (Cetoni) flow against measured hydrodynamic resistance.

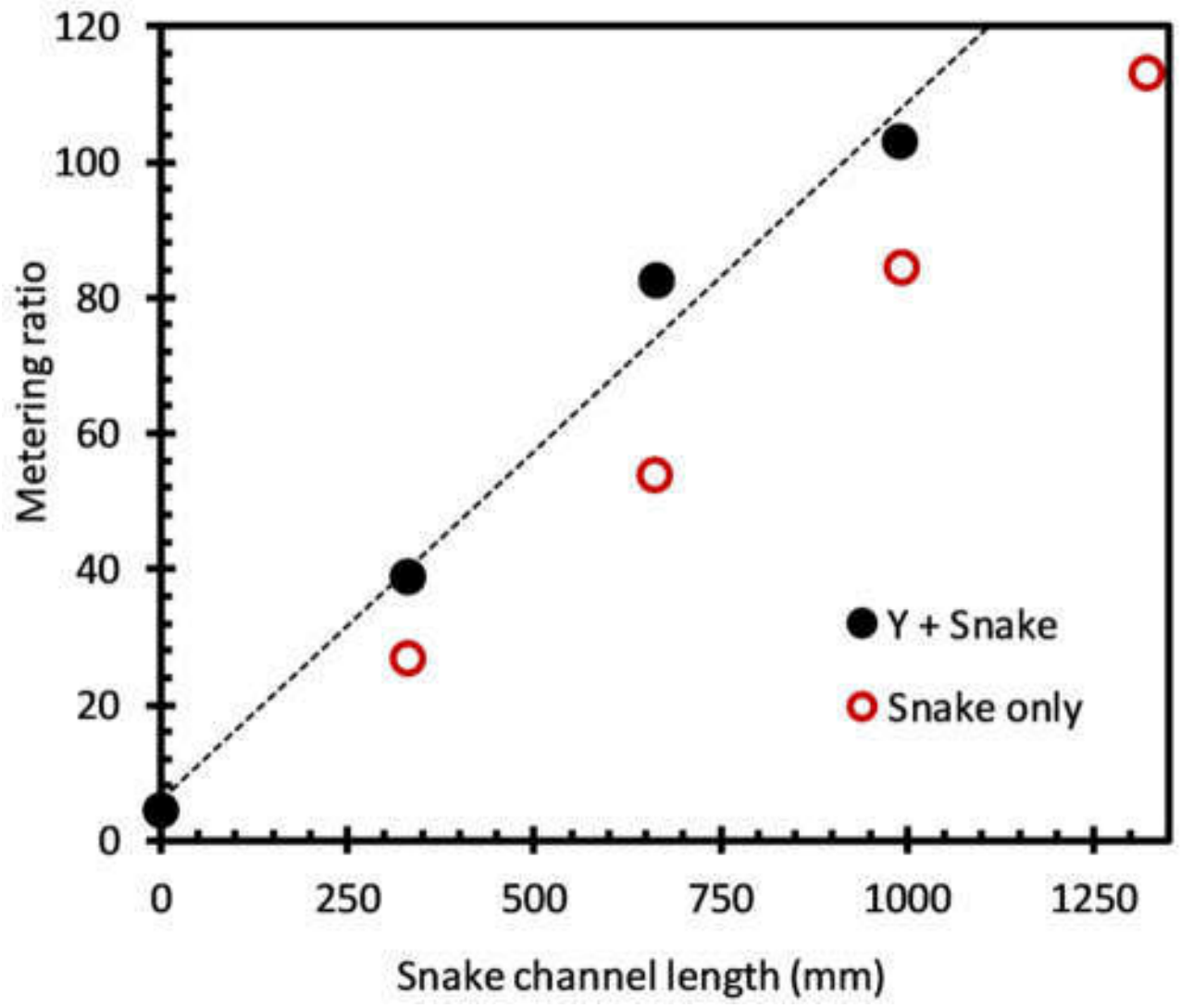

Fig. 20. Meter ratios for single snake channel with asymmetrical microfluidic Y-junction circuit, in comparison with those for snake channel alone.

Meter ratio variability was determined for each of the Y-meter devices, given in Fig. 19. The rms precision values were found to be in the range 0.15% - 0.21% while the solution conductivity / conductometer precision was ~0.10%, Fig. 21. These values are negligible with respect to impact on the overall $TCO_2$ measurement precision, Fig. 2.

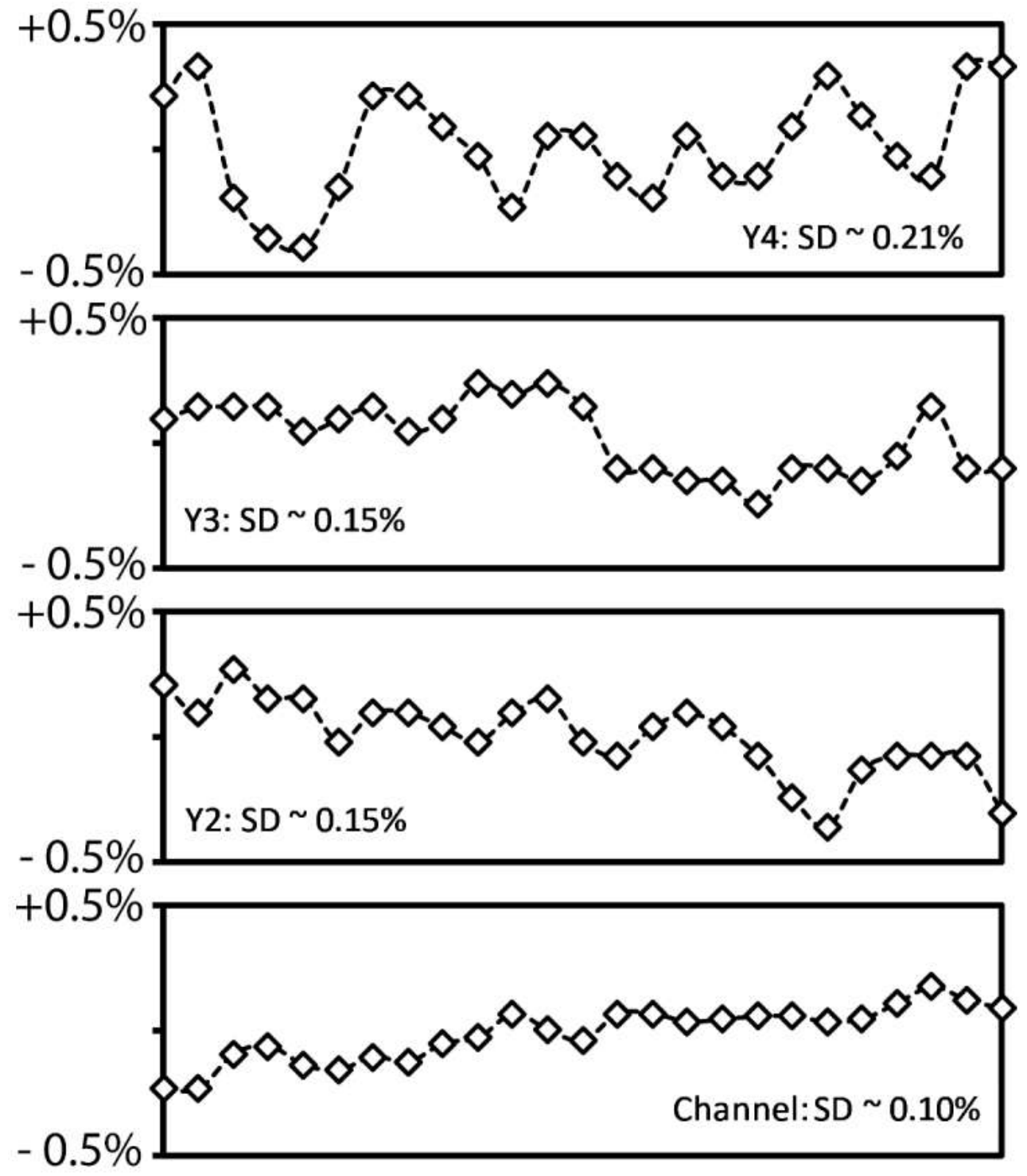

Fig. 21. Relative conductivity variation for 3 different Y meters, with high conductivity solution (1413 µS $cm^{-1}$) in high resistance line, and low conductivity solution (87 µS $cm^{-1}$) in low resistance line. Also shown is conductivity variation of channel only with direct dispense of KCl solution (84 µS $cm^{-1}$) without Y-meter.

## 5. Conclusion

Continuous and autonomous measurement of total dissolved inorganic carbon ($TCO_2$) in the oceans is critical for climate change studies, requiring miniaturised chemical analysis systems for ocean float-based deployment. A microfluidic conductivity-based approach offers the potential for such miniaturisation and sample pH reduction by acid injection to liberate dissolved $CO_2$ is required, which impacts on total device payload. To address this, we have fabricated laser etched microfluidic snake channel restrictors and asymmetric Y meters to adjust the metering ratio between sample and acid reagent. Laser etching conditions were varied to create a range of channel dimensions down to ~75 µm. Channel flow versus pressure measurements were used to determine hydrodynamic resistances which were compared with finite element simulations using a range of cross-section profiles and areas. Standard (laser raster mode) channels displayed specific $R_H$ values of 8 x $10^{12}$ Pa.s $m^{-3}$, while narrow/shallower channels (laser vector mode) were 140 – 300 x $10^{12}$ Pa.s $m^{-3}$. Microfluidic metering circuits were constructed from variable resistance snake channels and dimensionally symmetric or asymmetric Y-junctions. Sample to acid volume ratios (meter ratio) up to 100:1 have been achieved with 300 µm wide snake channel for lengths > 1m. At the highest pattern resolution, this would require a footprint of > 600 $mm^2$ (6 x$10^{-4}$ $m^2$). Circuits based solely on asymmetric Y-junctions gave meter ratios up to 16:1 with a footprint cost of < 40 $mm^2$ and precision values of ~0.2%. Meter ratio variability was negligible with respect to overall $TCO_2$ analysis, compared to the impact of temperature via reagent and sample viscosity changes. Temperature dependence can be negated with the use of high MR ratios above 75. Further design and fabrication refinements will be required to ensure the structural and dimensional integrity of such small channels in future integration of metering units into full $TCO_2$ analysis microfluidic circuits.

## Acknowledgements

The authors would like to acknowledge the funding support from: Invest N. Ireland (RD0714186), The Department of Employment and Learning, N. Ireland (US-IRL 013), Science Foundation Ireland (09/US/I1758), and National Science Foundation (US) (NSF 0961250).

We would also like to acknowledge the support and advice of Todd Martz and Phil Bresnahan of Scripps Institute of Oceanography, UC San Diego.

## References.